\documentclass[conference,a4paper]{IEEEtran}
\IEEEoverridecommandlockouts

\usepackage{fancyhdr}
\usepackage{graphicx}
\usepackage{indent}
\usepackage{amsmath,amssymb}
\usepackage{paralist}
\usepackage{eclbkbox}
\usepackage{subfigure}
\usepackage{multirow}

\makeatletter
\def\tbcaption{\def\@captype{table}\caption}
\def\figcaption{\def\@captype{figure}\caption}
\makeatother

\begin{document}
\title{Early Discovery of Chronic Non-attenders \\by Using NFC Attendance Management System
\thanks{\copyright 2013 IEEE. Personal use of this material is permitted. Permission from IEEE must be obtained for all other uses, in any current or future media, including reprinting/republishing this material for advertising or promotional purposes, creating new collective works, for resale or redistribution to servers or lists, or reuse of any copyrighted component of this work in other works.}
}

\author{\IEEEauthorblockN{Takumi Ichimura}
\IEEEauthorblockA{Faculty of Management and Information Systems,\\
Prefectural University of Hiroshima\\
1-1-71, Ujina-Higashi, Minami-ku,\\
Hiroshima, 734-8559, Japan\\
Email: ichimura@pu-hiroshima.ac.jp}
\and
\IEEEauthorblockN{Shin Kamada}
\IEEEauthorblockA{Graduate School of Comprehensive Scientific Research,\\
Prefectural University of Hiroshima\\
1-1-71, Ujina-Higashi, Minami-ku,\\
Hiroshima, 734-8559, Japan\\
Email: shinkamada46@gmail.com}
}

\maketitle

\fancypagestyle{plain}{
\fancyhf{}	% clear all header and footer fields
\fancyfoot[L]{}
\fancyfoot[C]{}
\fancyfoot[R]{}
\renewcommand{\headrulewidth}{0pt}
\renewcommand{\footrulewidth}{0pt}
}

\pagestyle{fancy}{
\fancyhf{}
\fancyfoot[R]{}}
\renewcommand{\headrulewidth}{0pt}
\renewcommand{\footrulewidth}{0pt}

\begin{abstract}
Near Field Communication (NFC) standards cover communications protocols and data exchange formats. They are based on existing radio-frequency identification (RFID) standards. In Japan, Felica card is a popular way to identify the unique ID. Recently, the attendance management system (AMS) with RFID technology has been developed as a part of Smart University, which is the educational infrastructure using high technologies, such as ICT. However, the reader/writer for Felica is too expensive to build the AMS. NFC technology includes not only Felica but other type of IC chips. The Android OS 2.3 and the later can provide access to NFC functionality. Therefore, we developed AMS for university with NFC on Nexus 7. Because Nexus 7 is a low cost smart tablet, a teacher can determine to use familiarly. Especially, this paper describes the method of early discovery for chronic non-attenders by using the AMS system on 2 or more Nexus 7 which is connected each other via peer-to-peer communication. The attendance situation collected from different Nexus 7 is merged into a SQLite file and then, the document is reported to operate with the trunk system in educational affairs section.
\end{abstract}


\IEEEpeerreviewmaketitle

\section{Introduction}
\label{sec:Introduction}
Regular attendance is an important factor in school success. Students who are chronic non-attenders receive fewer hours of instruction and often leave education early. It has been suggested that chronic absenteeism is not a cause of academic failure and departure from formal education, but rather one of many symptoms of alienation from school\cite{Rothman01}. Chronic absenteeism, truancy and academic failure may be evidence of a dysfunctional relationship between student and school. According to the change of social environment such as more relaxed education policy in Ministry of Education, Culture, Sports, Science, and Technology, it is required that university needs to be more student-centered and supportive of students with different needs.

Recently, the attendance management system (AMS) with RFID technology has been developed as a part of Smart University\cite{SmartCities}, which develops the educational infrastructure using high technologies, such as ICT. However, the reader/writer for Felica is too expensive to build the AMS. Furthermore, the reader/writer is set at each lecture room and the teacher cannot witness the student's authentication of the ID card, so that, one of friends can answer the roll call for an absentee.

Near Field Communication (NFC) standards cover communications protocols and data exchange formats. They are based on existing radio-frequency identification (RFID) standards. In Japan, Felica card is a popular way to identify the unique ID. NFC technology includes not only Felica but other type of IC chips. The Android OS 2.3 and the later can provide access to NFC functionality and some applications have been developed \cite{touchATND}. Therefore, we developed AMS for university with NFC on Nexus 7 \cite{Ichimura12, Kamada12, Ichimura13, NFCMovie, ITProducts}. Because Nexus 7 is a low cost smart tablet, a teacher can determine to use familiarly. Any teacher, who calls the roll at the class, can free from stress before the lecture. Especially, this paper describes the method of early discovery for chronic non-attenders by using the AMS system on 2 or more Nexus 7 which is connected each other via peer-to-peer communication. The attendance situation collected from different Nexus 7 is merged into a SQLite file and then, the document is reported to operate with the trunk system in educational affairs section.

Section \ref{sec:RFID} explains the Radio Frequency Identification. Section \ref{sec:NearFieldCommunication} describes NFC specification. In Section \ref{sec:NFC_AttendSystem}, the functions of our developed NFC Attendance Management System will be described. Some experiments for the discovery of serious absentees are investigated. In Section \ref{sec:ConclusiveDiscussion}, we give some discussions to conclude this paper.

\section{Radio Frequency IDentification}
Radio Frequency IDentification(RFID) is the use of a wireless no-contact system. The part of radio frequency uses electromagnetic fields to transfer data from the specified tag. Some tags require no electric power and are recognized at short ranges through electromagnetic induction. This section explains the basis of RFID to be easy to understand NFC (Near Field Communication) described in Section \ref{sec:NearFieldCommunication}.
 
\begin{figure*}[!tb]
\begin{center}
\includegraphics[scale=1.0]{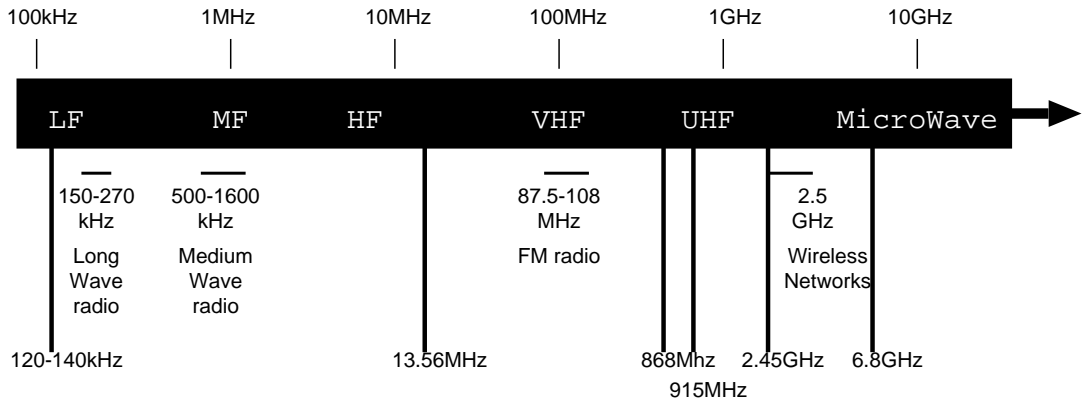}
\caption{Frequencies of RFID}
\label{fig:RFIDFrequencies}
\end{center}
\end{figure*}

\label{sec:RFID}
\subsection{RFID tags}
\label{sec:RFIDtags}
Table \ref{tab:ComparisonBarRFID} shows the comparison of bar codes and RFID Tags. The specifications of bar codes are as follows. They require line of sight to be read. They can only be read individually. They cannot be read if they become dirty or damaged. They must be visible to be logged. They can only identify the type of item and their information cannot be updated. They must manually tracked for item identification, making human error an issue. On the contrary, the specifications of RFID tags are as follows. They can be read or updated without line of sight. Multiple RFID tags can be read simultaneously. They are able to cope with harsh and dirty environments. They are very thin and can be printed on a label, and they can be read even if concealed within an item. They can identify a specific item. Electronic information can be over-written repeatedly on them. They can be automatically tracked, eliminating human error.

\begin{table}[tbp]
\begin{center}
\caption{The Comparison of Bar code and RFID Tags\cite{BarCodes}}
%\begin{tabular}{c|p{10zw}|p{10zw}}
\begin{tabular}{c|p{2.3cm}|p{2.3cm}}
\hline
                    & BAR CODES & RFID Tags \\ \hline
Technology          & Optical   & Radio frequency \\ \hline
Line of Sight       & Required  & Not required \\ \hline
Read Rates          & Slow - one at a time & Fast - up to 1,000 in a single pass \\ \hline
Read Range          & Inches to a few feet & Tens to hundreds of feet \\ \hline
Memory Capabilities & Static - read-only and limited data capacity & Dynamic - high capacity; reads, writes, updates, triggers other actions \\ \hline
Durability          & Exposed - risk of wear and tear or damage during handling & Better protected, can be encased, withstands harsh environments \\ \hline
Service Life        & Unlimited, subject to degradation & Up to 10 years \\ \hline
Security            & Low - easily copied or faked & High - encryption is harder to replicate \\ \hline
Interference        & May be subject to obstruction from dirt or damage from handling & Metal and liquids can interfere with some frequencies \\ \hline
Re-usability         & No & Yes \\ \hline
Cost                & Fractions of a penny to a few cents & Up to \$50 \\ \hline
Human Labor         & Required - high & For hand-held readers - moderate Not required for fixes readers \\ \hline
\end{tabular}
\label{tab:ComparisonBarRFID}
\end{center}
\end{table}

\subsection{RFID Frequencies}
\label{sec:FrequenciesRFID}
RFIDs are considered to be radio emitters, and therefore must follow a set of frequency regulations. There is no global public body that governs the frequencies used for RFID. The main bodies governing frequency allocation in Japan for RFID is Ministry of Internal Affairs and Communications.

Fig.\ref{fig:RFIDFrequencies} shows the frequencies of RFIDs\cite{RFidea}. As shown in Fig.\ref{fig:RFIDFrequencies}, LF(Low-frequency) is 125 - 134.2 kHz and 140 - 148.5 kHz and HF(high-frequency) is 13.56 MHz. RFID tags for LF and HF can be used globally without a license. UHF(Ultra-high-frequency) is 868 MHz-928 MHz. RFID tags for UHF cannot be used globally as there is no single global standard. In North America, UHF can be used unlicensed for 908 - 928 MHz, but restrictions exist for transmission power. In Europe, UHF can be used in the 865.6 - 867.6 MHz band. For Australia and New Zealand, 918 - 926 MHz are unlicensed, but restrictions exist for transmission power. These frequencies are known as the ISM bands (Industrial Scientific and Medical bands).

\begin{figure}[tb]
\begin{center}
\includegraphics[scale=0.8]{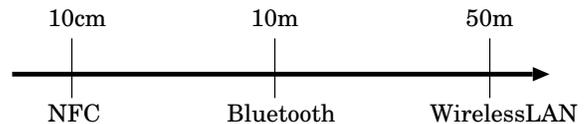}
\caption{Communication range}
\label{fig:NFCdistance}
\end{center}
\end{figure}

\subsection{Passive or Active RFID tags\cite{Technovelgy}}
\label{sec:PassiveActiveRFIDtags}
A passive tag is one of RFID tags that does not contain a battery, because the power is supplied by the reader. When radio waves from the reader are encountered by a passive RFID tag, the coiled antenna within the tag forms a magnetic field. The tag draws power from it, energizing the circuits in the tag. The tag then sends the information encoded in the tag's memory.

On the contrary, an active tag is one of RFID tags when it is equipped with a battery that can be used as a partial or complete source of power for the tag's circuitry and antenna. Some active tags contain replaceable batteries for years of use; others are sealed units.

The advantages of a passive tag are as follows.
\begin{itemize}
\item The passive tag functions without a battery.
\item The tag is typically much less expensive to manufacture.
\item The tag is much smaller. These tags have almost unlimited applications in consumer goods and other areas.
\end{itemize}

\section{Near Field Communication}
\label{sec:NearFieldCommunication}
\subsection{Range and Distance}
\label{sec:NFCtag}
Recently, NFC (Near Field Communication) in the RFID tags has attracted a great deal of engineers attention as shown in Table \ref{tab:ComparisonBarRFID} and Subsections \ref{sec:FrequenciesRFID} and \ref{sec:PassiveActiveRFIDtags}. NFC is a wireless communication technology operating at 13.56 MHz over a short distance of about maximum 10 centimeters. This technology enables communication among electronic devices brought into close range of each other, as well as between such devices and conventional contact-less IC cards\cite{SONY}.

Fig.\ref{fig:NFCdistance} shows the communication range of computer network used in daily life; NFC, Bluetooth, and Wireless LAN. 

\subsection{NFC cards}
\label{sec:NFCcards}
NFC is derived from communication technology which is specified by the International standard ISO/IEC 18092 (NFCIP-1). The NFC Forum\cite{NFCForum} provides a highly stable framework for extensive application development, seamless interoperable solutions, and security for NFC-enabled transactions. The NFC Forum has organized the efforts of dozens of member organizations by creating Committees and Working Groups.

\begin{figure}[tb]
\begin{center}
\includegraphics[scale=0.8]{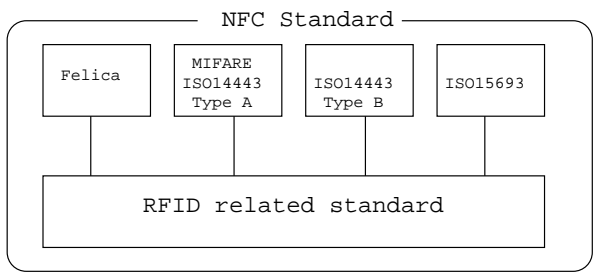}
\caption{NFC standard}
\label{fig:NFCstandard}
\end{center}
\end{figure}

\begin{figure}[tb]
\begin{center}
\includegraphics[scale=0.45]{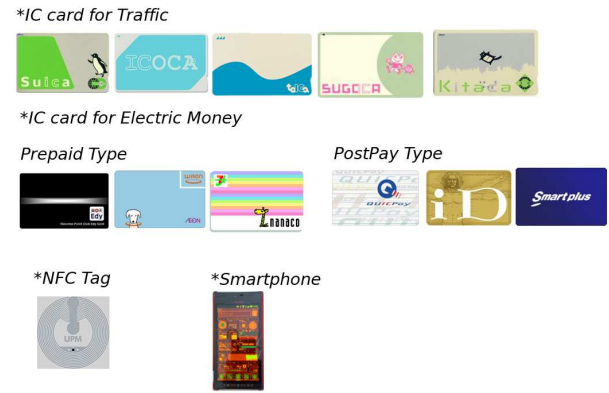}
\caption{Example of RFID cards}
\label{fig:ICcards}
\end{center}
\end{figure}

Fig.\ref{fig:NFCstandard} shows the concept of NFC. In the NFC Forum specifications, the Type-A and Type-B communication technologies specified in the contact-less IC card international standard ISO/IEC 14443. Their cards called NFC-A and NFC-B, respectively. The Felica communication technology is called NFC-F and the Japanese Industrial Standard JIS X 6319.4. Fig.\ref{fig:ICcards} shows the example IC cards which we can use in Japanese.

A standards-based connectivity technology, NFC harmonizes today's diverse contact-less technologies, enabling current and future solutions in areas such as:
\begin{itemize}
\item Access control
\item Consumer electronics
\item Healthcare
\item Information collection and exchange
\item Loyalty and coupons
\item Payments
\item Transport
\end{itemize}
However, the contact-less IC card and its related products have been limited to the cards themselves and reader/writer.

\subsection{NFC mode}
\label{sec:NFCmode}
NFC Forum specifications are based on existing and recognized standards like ISO/IEC 18092 and ISO/IEC 14443-2,3,4, as well as JIS X6319-4\cite{NFCForum}. The specifications describe the parts of those standards that are relevant for NFC Forum devices. Currently, certification includes testing for the lower-level digital protocols, specifically the Tag Operation specifications for the different tag types, the NFC Digital Protocol Specification, and the NFC Activity Specification. Moving forward, certification will also include testing for the physical layer ``NFC RF Analogue Technical Specification'', and selected upper-level digital protocols: ``NFC Logical Link Control Protocol (LLCP) Technical Specification'' and ``NFC Simple NDEF Exchange Protocol (SNEP) Technical Specification''.

\begin{figure}[tb]
\begin{center}
\includegraphics[scale=0.7]{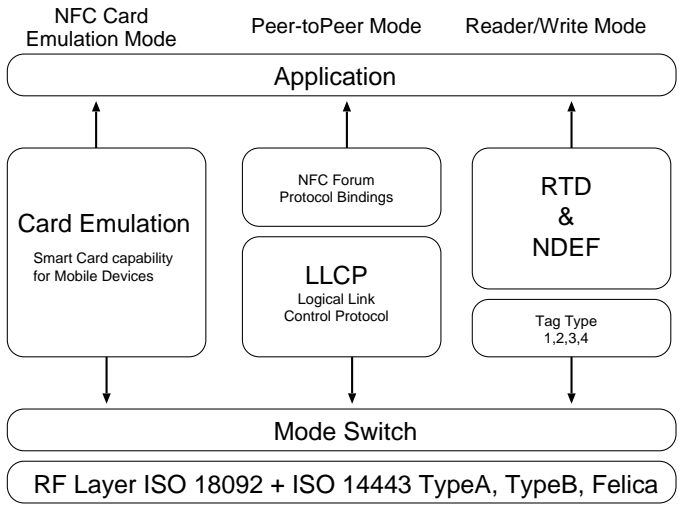}
\caption{NFC mode}
\label{fig:NFCmode}
\end{center}
\end{figure}

Fig.\ref{fig:NFCmode} shows the 3 modes of NFC: Card Emulation mode, Peer to Peer mode, and Reader/Writer mode. An NFC device in card emulation mode can replace a contact-less card or tag. This will enable NFC devices to be used with existing contact-less card infrastructure in applications such as ticketing, access control and payments.

Peer to peer mode enables two NFC devices to share data between them. Due to the low transfer speed of NFC if large amounts of data need to be sent, peer to peer mode can be used to create a secondary high speed connection (handover) like Bluetooth or WiFi. In this case, the NFC is used to negotiate the optimal communication protocol and transfer authentication data for the secondary protocol. The file or data is then sent over the high capacity protocol (i.e. Bluetooth, WiFi, etc).

NFC enabled devices are able to read and write to NFC and many contact-less cards. For example, if a NFC tag is attached to a poster, the NFC smartphone can "tap" the tag to access the information stored in the tag (e.g. coupons, maps, product information, etc) easy and conveniently.

\section{NFC Attendance Management System}
\label{sec:NFC_AttendSystem}
In order to be more student-centered and supportive of students with different needs, the project of Smart University \cite{SmartCities} is underway in some universities. Smart University is inspired by Smart City that the concept of the smart city has been introduced as a strategic device to encompass modern urban production factors in a common framework and to highlight the growing importance of Information and Communication Technologies (ICTs), social and environmental capital in profiling the competitiveness of cities.

Recently, the attendance management system (AMS) with RFID technology has been developed as a part of Smart University \cite{SmartCities}, which develops the educational infrastructure using high technologies, such as ICT. However, the reader/writer for Felica is too expensive to build the AMS. Furthermore, the reader/writer is set at each lecture room and the teacher cannot witness the student's authentication of the ID card, so that, one of friends can answer the roll call for an absentee.

NFC reader/writer technology is equipped with the smartphone or smart tablet such as Google Nexus 7. The smart tablet can easily be carried around, so the student must be checked the attendance in the teacher's presence. Therefore, the records in the system are correctly and the student become an earnest student to attend the lecture for real.

\begin{figure}[tb]
\begin{center}
\includegraphics[scale=0.40]{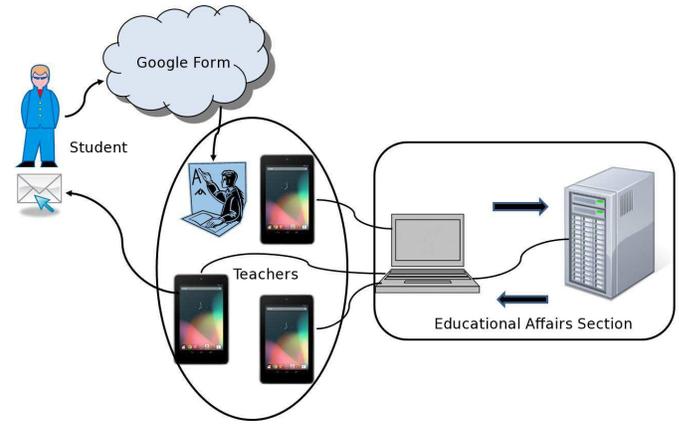}
\caption{System Overview}
\label{fig:SystemOverview}
\end{center}
\end{figure}

Fig.\ref{fig:SystemOverview} shows the system overview of our developed AMS system and the system of educational affair's section. The Nexus part is used to call the roll by teachers. The educational affairs section collects the data related to the absentee by connecting the Nexus to PC after the lecture. The data communication is via Bluetooth connection or the directed connection with microUSB. If the IC card reader/writer is equipped with the PC, data is transmitted automatically after the peer-to-peer connection with NFC authentication.

\subsection{NFC AMS on smart tablet \cite{Ichimura12, Kamada12, Ichimura13, NFCMovie, ITProducts}}
This section describes the functions of NFC AMS system on smart tablet. The system is all-in-one software which has functions to register the information class in charge, to make a list of students, to take attendance, and create follow-up personal message and send it one by one. The teacher can see the reason of the absence through Google Drive.

\begin{figure}[tbp]
\begin{center}
\subfigure[Student List]{
\includegraphics[scale=0.24]{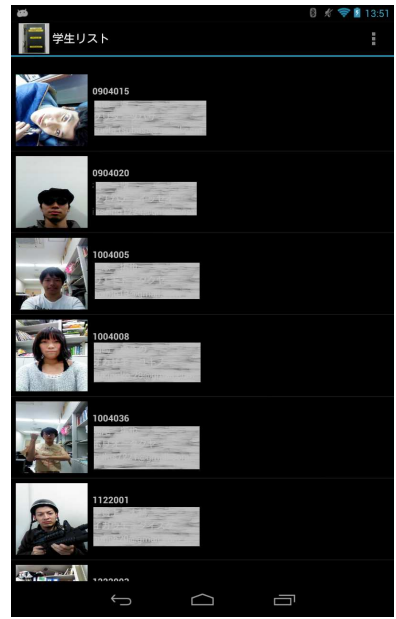}
\label{fig:face0}
}
\subfigure[Take a picture]{
\includegraphics[scale=1.25]{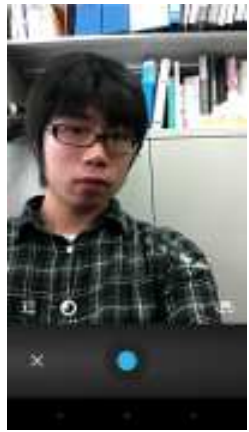}
\label{fig:face1}
}
\subfigure[Normal Attendance]{
\includegraphics[scale=0.27]{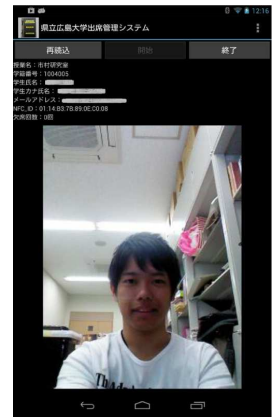}
\label{fig:normalattend}
}
\caption{Alert Display}
\label{fig:Alerts0}
\end{center}
\end{figure}

\begin{figure}[tbp]
\begin{center}
\subfigure[Alert(two or more consecutive absents)(Yellow)]{
\includegraphics[scale=1.5]{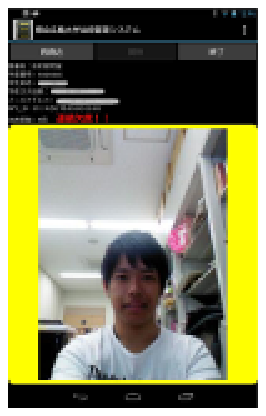}
\label{fig:alertserialattend}
}
\subfigure[Alert(Many Absents)(Yellow)]{
\includegraphics[scale=1.5]{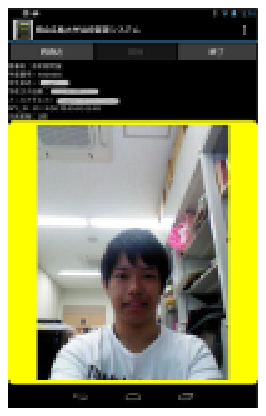}
\label{fig:alertyellowattend}
}
\subfigure[No Accreditation(Red)]{
\includegraphics[scale=1.5]{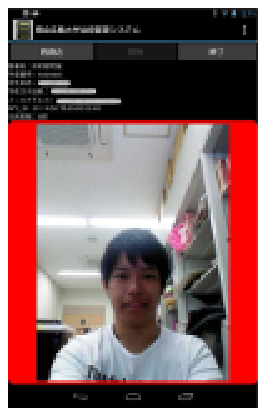}
\label{fig:alertredattend}
}
\caption{Alert Display}
\label{fig:Alerts1}
\vspace{-3mm}
\end{center}
\end{figure}

\begin{itemize}
\item Preparation of class information\\
The teacher selects the specified information of class from educational affair's section system and makes a csv file which includes student ID, Name and email address. The NFC AMS inserts the csv file into SQLite database management system.
\item Register NFC UID of students IC card\\
Because Nexus 7 is equipped with NFC reader/writer, it can read the UID (Unique ID) of NFC card or IDm of Felica card. The user touches the line of student to register NFC card and a face picture student list as shown in Fig.\ref{fig:face0}, and the camera application as shown in Fig.\ref{fig:face1} starts. The number of UID or IDm is scanned and the face of the student is displayed automatically as shown in Fig.\ref{fig:normalattend}, if the card closes to the Nexus 7. Then, the teacher can recognize the attendance of  corresponding student actually.
\item Alert to chronic non-attenders\\
When the NFC card closes to the Nexus 7, the system informs to the student who has many absences or is continuation absence, by beep sound or colored display as shown in Fig.\ref{fig:alertserialattend}, Fig.\ref{fig:alertyellowattend}. Fig.\ref{fig:alertredattend} shows the situation of not-approval of credits for the class by regulations.
\item Follow-up email\\
Follow-up email is sent automatically to the absentee at the class when the teacher closes the system. The message has the class name, student name, the number of absences, and the specified URL, which is generated individually. The URL is Google form and includes the information the class name, student ID, date of lecture and so on as shown in Fig.\ref{fig:GoogleForm}. The student checks the follow-up email and click the URL, and then he/she can make a confirmation of absence and send the reason for his/her absence and the message to teacher. The teacher can recognize whether the absence of the student is truancy. 
\item Tabulation on Nexus
Fig.\ref{fig:Tabulation} shows the tabulation of attendance list on Nexus. The function is effective for the use in class, if the student protests the attendance.
\item Peer-to-peer data communication\\
After the authentication by NFC, the Peer-to-peer data communication starts. The AMS data such as the records of attendance on 2 or more Nexus 7s are merged to be complementary to each other. The function is effective to use in large classroom, because 2 ore more Nexus 7s can be used in a class and the records are synchronized each other.
\item Backup and Restore
The special edition of NFC AMS system can make a backup file in SD card. Of course, it can restore from the data dump. 
\item Handling instructions\\
The detailed explanation of the system can see the Youtube movie in the system for handling instructions\cite{NFCMovie}.
\end{itemize}

\subsection{Operation with educational affair’s section}
\begin{figure}[tbp]
\begin{center}
\subfigure[Google Form for absentee]{
\includegraphics[scale=0.3]{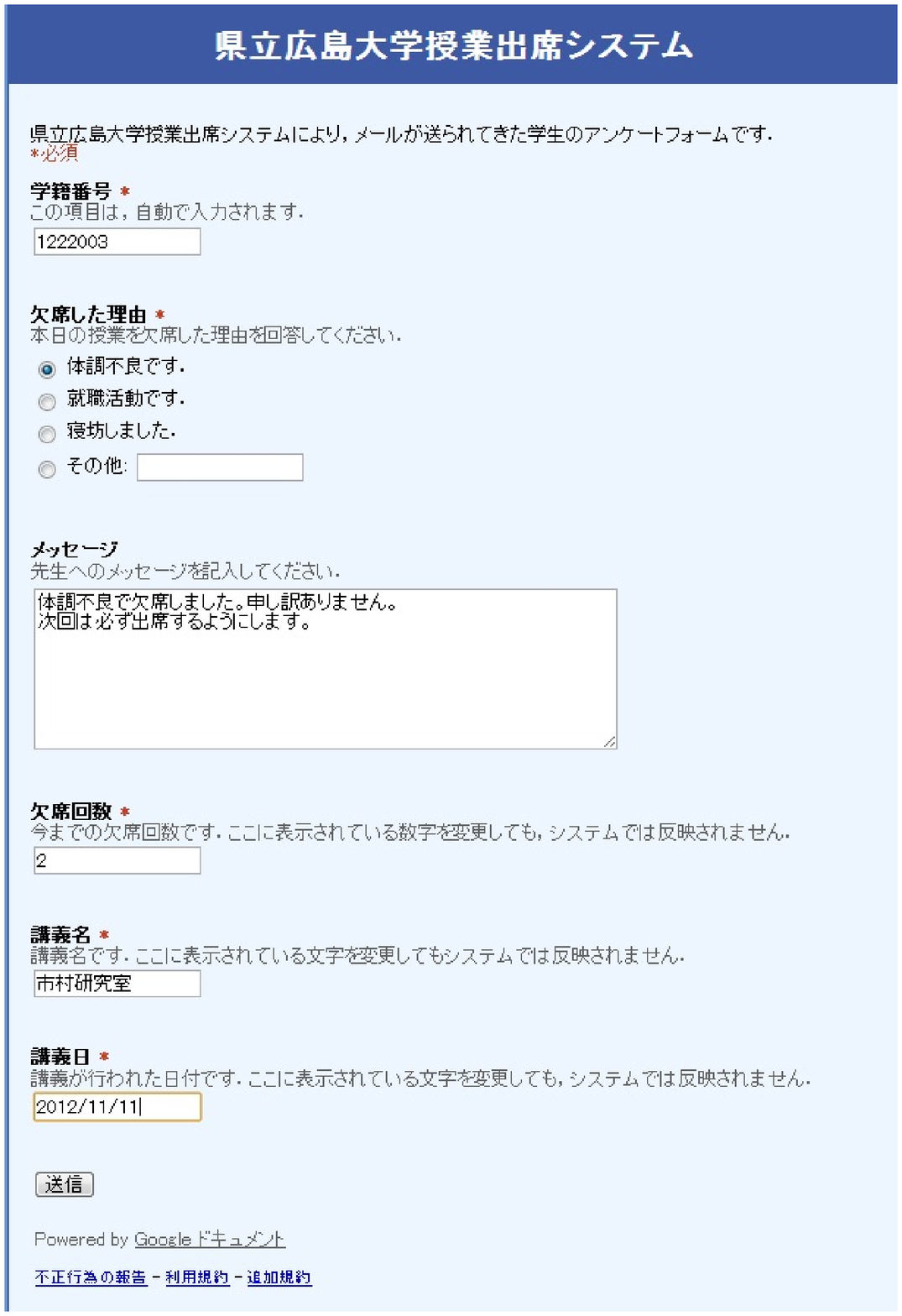}
\label{fig:GoogleForm}
}
\subfigure[Tabulation on NFC AMS]{
\includegraphics[scale=0.18]{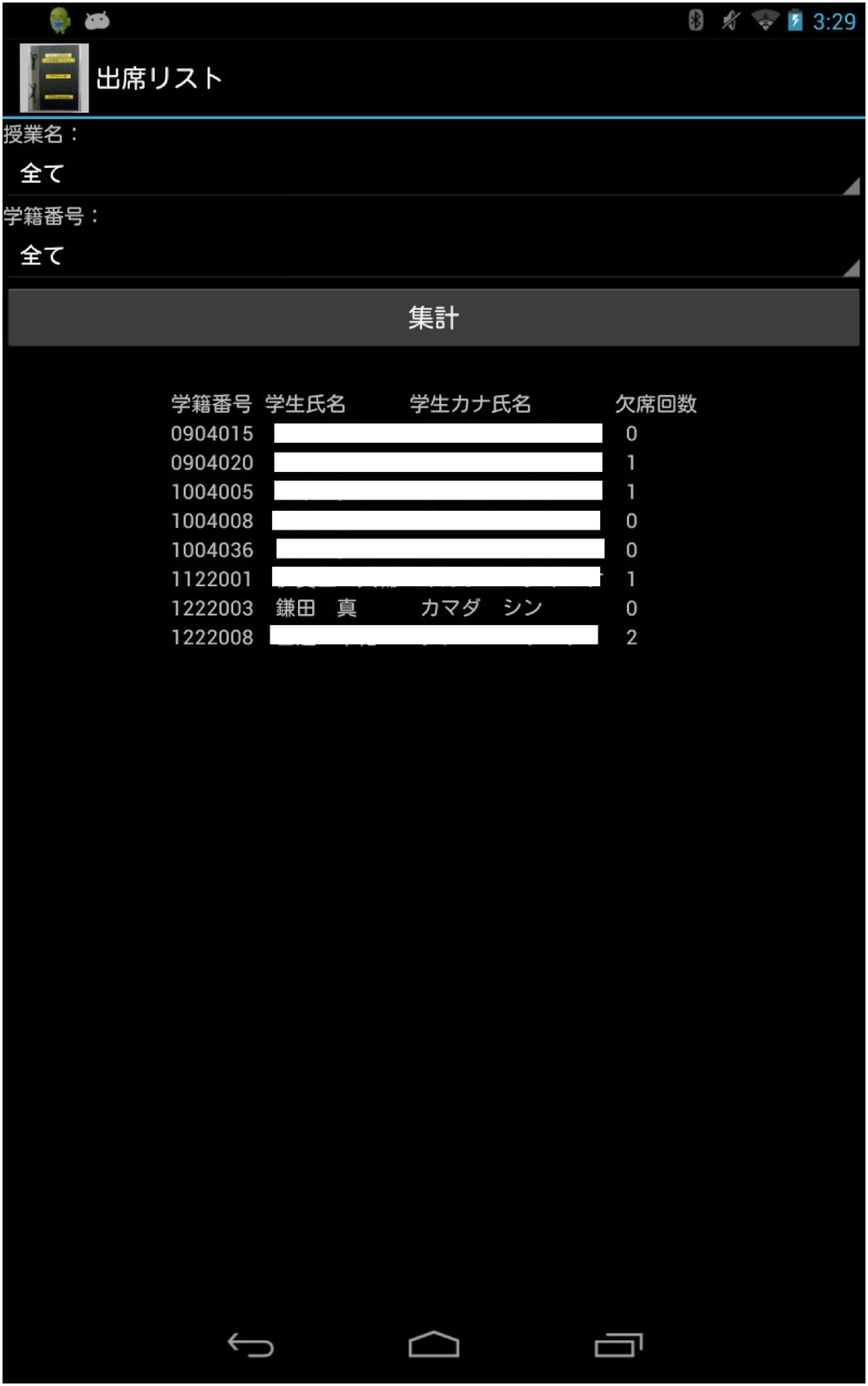}
\label{fig:Tabulation}
}
\subfigure[SQLite Browser on Firefox]{
\includegraphics[scale=0.3]{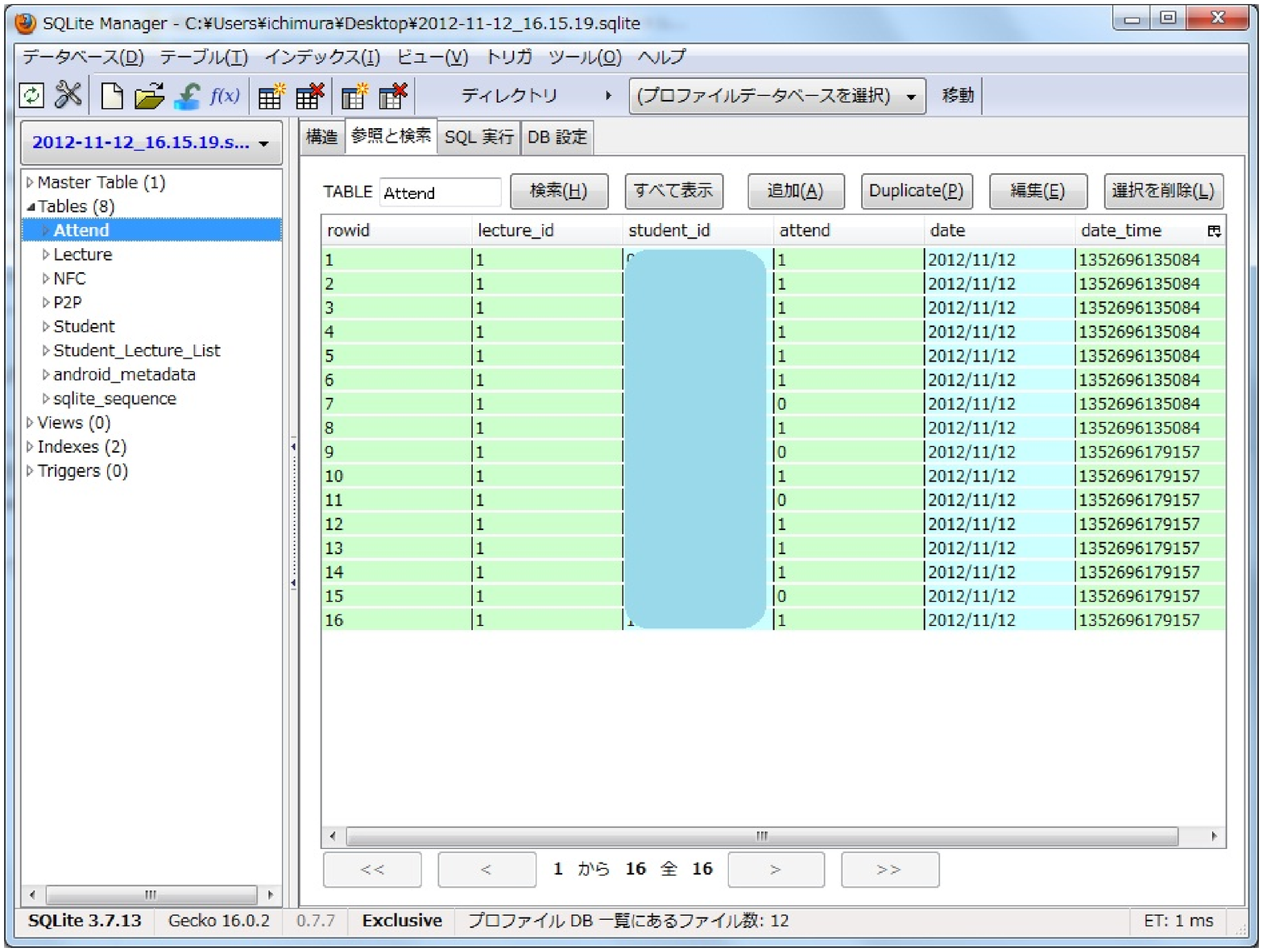}
\label{fig:SqliteBrowser}
}
\caption{Alert Display}
\label{fig:Alerts}
\vspace{-3mm}
\end{center}
\end{figure}

Fig.\ref{fig:SqliteBrowser} shows SQLite database management system which works in Firefox on PC. The SQLite is a software library that implements a self-contained, server less, zero-configuration, transactional SQL database engine. Therefore, the user can use the sql command such as `select', `insert', `update', and so on. For the operation in educational affair' section, some SQL command as shown in Fig.\ref{fig:sql} are prepared to select the specified data of students.

\begin{center}
%\begin{indentation}{0.3zw}{0.3zw}
\begin{indentation}{0.2cm}{0.2cm}
\begin{breakbox}
\smallskip
{\begingroup
\baselineskip=13pt \lineskiplimit=-\maxdimen
\begin{verbatim}
SELECT Attend.student_id , name1 , name2,
SUM(CASE WHEN attend = '0' THEN 1 ELSE 0 
END) AS 'Absent' FROM Attend , Student 
WHERE Attend.student_id=Student.student
_id AND lecture_id='1' GROUP BY 
Attend.student_id  Having SUM(CASE WHEN 
attend='0' THEN 1 ELSE 0 END) >=1 ORDER 
BY Attend.student_id;
\end{verbatim}
\endgroup}
\end{breakbox}
\end{indentation}
\vspace{-2mm}
\figcaption{SQL Example}
\label{fig:sql}
\end{center}

\section{Conclusion}
\label{sec:ConclusiveDiscussion}
Chronic absenteeism, truancy and academic failure may be evidence of a dysfunctional relationship between student and school. According to the change of social environment, it is required that university needs to be more student-centered and supportive of students with different needs. The NFC Attendance Management System can check the student's attendance. Moreover, the teacher can form a relationship between teacher and student via the system, because the teacher can recognize the faces displayed on Nexus 7 when the student touch the ID card to it. Although many investigations are not implemented, absentees at almost classes decreased. We consider that the student has the consciousness of being observed. The use in some classes in 2 or more universities will be considered for the operation test in near future.

\end{document}